%% file: sln-fund-npb.tex
\documentclass[a4paper,showpacs,showkeys,pdf]{revtex4}
\usepackage{young}
\usepackage{amsfonts}
\usepackage{amssymb}
\usepackage{amsmath}
\usepackage{amsthm}
\usepackage{revsymb}
\usepackage{youngtab}
\usepackage{epic}
\usepackage{eepic}
\newcommand{\be}{\begin{equation}}
\newcommand{\ee}{\end{equation}}
\newcommand{\bea}{\begin{eqnarray}}
\newcommand{\eea}{\end{eqnarray}}

\newcommand{\com}[2]{\left[#1,#2\right]}

\newcommand{\sln}{\mathfrak{sl}(n,\mathbb{C})}
\newcommand{\sun}{{\mathfrak{su}(n)}}
\newcommand{\sll}[1]{\mathfrak{sl}(#1,\mathbb{C})}
\newcommand{\su}[1]{\mathfrak{su}(#1)}
\newcommand{\cartan}{\mathfrak{h}}
\newcommand{\borel}{{\mathfrak{b}}}

\newcommand{\ind}[1]{{\boldsymbol{#1}}}
\newcommand{\fund}{\text{f}}

\newcommand{\ham}{\mathcal{H}}
\newcommand{\hamt}{\widehat{\mathcal{H}}}
\newcommand{\en}{\mathcal{E}}

\newcommand{\sptot}{{\mathbb V}^N}

\newcommand{\sptotN}[1]{{\mathbb V}^#1}

\newcommand{\spin}{\boldsymbol{S}}
\newcommand{\eps}{\varepsilon}

\newcommand{\yn}{\mathbb{Y}}
\newcommand{\yc}{\Yvcentermath1}

\newcommand{\rep}{\rho}
\newcommand{\bras}[1]{|#1\rangle}

\newcommand{\bra}[1]{\left|#1\right\rangle}
\newcommand{\ket}[1]{\left\langle #1\right|}

\newcommand{\brav}[1]{\overline{\bra{#1}}}

\newcommand{\gs}{\Omega}
\newcommand{\perm}{P}
\newcommand{\gperm}{{\rm Perm}}

\newtheorem*{theorem*}{Theorem}
\newtheorem{theorem}{Theorem}%[section]
\newtheorem{corollary}[theorem]{Corollary}
\newtheorem{lemma}[theorem]{Lemma}
\newtheorem{proposition}[theorem]{Proposition}
\newtheorem*{proposition*}{Proposition}
\theoremstyle{definition}
\newtheorem{definition}{Definition}%[section]
\theoremstyle{remark}
\newtheorem{remark}{Remark}%[section]
\newtheorem*{notation}{Notation}
\newtheorem{example}{Example}%[section]

\newcommand{\bs}{\baselineskip}

\newcommand{\seq}[2]{\{#1,\dots,#2\}}

\DeclareMathOperator{\tr}{Tr}

\newcommand{\asym}{\mathcal{Q}^-}

\DeclareMathOperator{\id}{\openone}
\DeclareMathOperator{\spann}{span}
\DeclareMathSymbol{\C}{\mathalpha}{AMSb}{"43}

\begin{document}

%\preprint{YerPhI-1577(2)-2002}

\title{The ordering of energy levels for $\sun$ symmetric antiferromagnetic chains}

\author{Tigran Hakobyan}

\email{hakob@lx2.yerphi.am}

\affiliation{Theory Division, Yerevan Physics Institute,
    Alikhanyan Street 2, 375036 Yerevan, Armenia}
\date{\today}

%\pacs{75.10.Jm 75.50.Ee 71.10.-w 68.35.Rh 05.50.+q 02.30.Ik}



\begin{abstract}
The $\sun$ symmetric antiferromagnetic finite chain with the fundamental
representation and nearest-neighbor interaction is studied.
A partial ordering between the lowest energy levels $\en(\yn)$ in multiplet
sectors corresponding to different Young tableaux $\yn$ is established
for the chains with arbitrary site-dependent couplings.

For the open chains it is proved that $\en(\yn_1)>\en(\yn_2)$ if $\yn_2$
may be obtained from $\yn_1$ moving down some of its boxes. In particular,
the ground state of the chain belongs to the antisymmetric multiplet
sector. For the rings the same condition is fulfilled if, in addition, all rows
of $\yn_2$ are of even or odd length. The ground state is $\sun$ singlet
if the chain's length is a multiple of $n$ both for open and periodic
chains.

The results generalize the well known Lieb-Mattis theorem to chains with
higher symmetries.
\end{abstract}

\maketitle

\section{Introduction}
%\label{sec:intro}

The properties of the ground state of one-dimensional spin systems
have been studied intensively over a long period. In most cases
the numerical simulations and perturbative methods are only
investigation tools. Analytic solutions to these systems are
rarely available therefore any exact result in this area is highly
appreciated. Among them we mention the exact solution of the spin
$S=1/2$ nearest-neighbor antiferromagnetic Heisenberg chain by
Bethe \cite{B31}. Later Bethe ansatz was applied to some other
one-dimensional spin models. In particular, nearest-neighbor
$\sun$ chain with spins living in $n$-dimensional fundamental
representation was solved exactly by Sutherland \cite{Suth}. Since
the original works \cite{MG,AKLT}, much attention have been paid
also to those spin models, which are not integrable but whose
ground states can be constructed exactly.

Nevertheless in many cases it is possible somehow to retrieve information
about the ground state without possessing the exact solution.
It was proved rigorously that the ground state of antiferromagnetic Heisenberg
$S=1/2$ ring with an even number of sites has total spin zero,
i.e. is a singlet \cite{M55}. This result was generalized
to higher spins and dimensions \cite{LSM61}.
The uniqueness of ground state had been proved also
\cite{LSM61,LM62,Lieb89}. The approach is extended later to $\sun$ chains
with self-conjugate antisymmetric representation \cite{AL86}.
The major advantage of this method is that it can be applied to rather general class
of models with \emph{arbitrary} antiferromagnetic spin exchange couplings, depending on
lattice site, and defined on any \emph{bipartite} lattice \cite{LM62,AL86}.
Note however that recently the exact results have been obtained
for frustrated spin systems on lattices with reflection symmetry
without demanding their bipartiteness \cite{LS99,LS00}.

For the antiferromagnet on bipartite lattice with rather
general $\su{2}$ Hamiltonian  Lieb and Mattis \cite{LM62,Lieb89}
proved exactly that among all spin-$S$ multiplets only one exists with minimal
eigenvalue $\en(S)$.
For the chains with open boundary conditions and for the rings with even number
of spins, the lowest energy level $\en(S)$ is ordered in a natural way, i.e.
it is an increasing function of the spin $S$.
In particular, the ground state of spin-$1/2$  open chain with even number of spins
is spin-singlet $S_\text{gs}=0$ and unique, while with odd number of spins it is
spin-doublet $S_\text{gs}=1/2$ and twofold degenerate.
For periodic chain the same condition is fulfilled provided that it has
an even number of spins.
Note that very recently the ordering of the energy levels for ferromagnetic $S=1/2$
Heisenberg chain have been established and proved \cite{NSS03}.

In this article we formulate and prove the analogue of this theorem
for $\sun$ symmetric open and periodic chains with nearest-neighbor interaction and
sites endowed with the fundamental representation.

The models with such symmetry have many applications in condensed matter physics.
In particular, the low energy properties of Sutherland chain
are described by conformal field theory \cite{Af86}.
Moreover, the spin systems, whose Hamiltonians contain higher order spin exchange terms,
may possess $\sun$ symmetry at the certain point of the phase space.
For instance, $\su{4}$ symmetric point exists in well known
spin-orbital model \cite{LMSZ98,YSU98} and two-leg spin ladder model \cite{W99}.
Recently $S=1$ Heisenberg model with additional biquadratic interactions
possessing $\su{3}$ symmetric point \cite{HK02,BOG02} have been investigated.
The systems with higher symmetries are used in
quantum computation and quantum information \cite{TV00}.

For the convenience,  the main result of the article is
given in the next section, while the subsequent sections are devoted
to its complete proof.

In Section~\ref{sec:main} a partial order for Young tableaux is introduced.
Then the main theorem of this article, which sets up a partial ordering of the lowest
energy levels of different sectors corresponding to different
Young tableaux, is formulated.

Section~\ref{sec:sln} contains some known results from the representation
theory of $\sln$ Lie algebra, which are used in the subsequent sections.

%The irreducible representations are
%classified by Young tableaux and can be decomposed on subspaces
%(called weight subspaces) having the same eigenvalues (called weights)
%under the action of the diagonal subalgebra of $\sln$.

In Section~\ref{sec:relative} the relative ground states on the weight
subspaces are studied.
For open chains they are proved to be unique, while for rings
the uniqueness is proved for the weight subspaces with
either even or odd color multiplicities.

In Section~\ref{sec:ordering} the multiplets, to which the relative ground
states belong, are determined.
%In particular, if the weight is a highest weight of some multiplet
%then the  corresponding relative ground state is the highest weight state
%of this multiplet.
The lowest energy levels in the sectors related to non-equivalent
multiplets  are compared to each other. It is proved that they are
ordered according to the partial order for corresponding Young tableaux,
introduced in Section \ref{sec:main}. This completes the proof of the
main theorem.

The results are summarized and discussed in Section~\ref{sec:conclusion}.

Appendix contain the proof of well known Perron-Frobenius
theorem about the non-degeneracy of the minimal eigenvalue of connected matrix
with non-positive off-diagonal elements.

\section{Main result}
\label{sec:main}

\begin{figure}
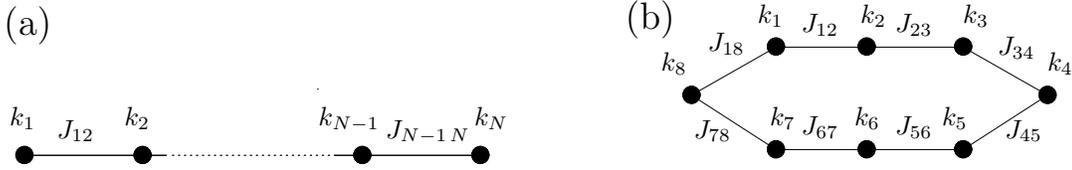

\begin{center}
\input chain.eepic
\end{center}
\caption{{\sf (a)}~The one-dimensional $N$-site Sutherlang chain
with open boundary conditions (open chain). {\sf (b)}~The $8$-site
chain with periodic boundary conditions (ring).} \label{fig:chain}
\end{figure}

The Hamiltonian of $\sun$ symmetric nearest-neighbor
antiferromagnetic $N$-site chain (see Fig.~\ref{fig:chain}) with
fundamental representation has the following form: \be
\begin{split}
  \label{H}
  \ham=\sum_{l=1}^{D(N)}\sum_{a=1}^{n^2-1} J_{l\,l+1}  X^a_\ind{l} X^a_\ind{l+1},
  \quad
  \text{where}
  \quad
  D(N)=
    \begin{cases}
    N-1 & \text{for open chains} \\
    N   & \text{for periodic chains}
    \end{cases}
    \quad
    \text{and \quad $J_{l\,l+1}>0$}.
\end{split}
\ee
Here $n^2-1= \dim \sun$ and the first and ($N+1$)-th sites of the periodic chain are identified.
The elements $iX^a$ are antihermitian matrices acting on $n$-dimensional space $V_n$
and forming the basic of $\sun$ Lie algebra in the fundamental representation.
The bold subscript in $X^a_\ind{l}$ means that the operator $X^a$ acts the space $V_n$
attached to $l$-th site of the chain.
The basic elements obey the commutation relations $\com{X^a}{X^b}=i\sum_c f_{abc} X^c$, where
$f_{abc}$ is a real antisymmetric tensor and are normalized as
$\tr\left(X^aX^b\right)=\delta_{ab}$.

If all couplings take equal values $J_{l\,l+1}=J$ then the Hamiltonian
\eqref{H} is reduced to well known Sutherland chain, which is exactly
solvable by Bethe ansatz \cite{Suth}.

%The equivalence classes of irreducible $\sun$ representations are
%characterized by different Young tableaux $\yn$ with row number not greater than $n$.

The total space of states of the chain \eqref{H} can be decomposed as
\be
\label{young-sp}
\sptot=\bigoplus_\yn \sptot_\yn, \qquad \text{where} \qquad \sptot:=\bigotimes_{i=1}^{N}V_n.
\ee
Here $\sptot_\yn$ is the subspace of $\sptot$ formed by all $\sun$ multiplets  belonging
to the \emph{same} equivalence class characterized by the Young tableau $\yn$.
We'll call it in the following a $\yn$ multiplet sector.
Due to $\sun$ symmetry, the
Hamiltonian $\ham$ has block-diagonal form with respect to this
decomposition.

The equivalence classes of irreducible $\sun$ representations, which can appear
in the decomposition \eqref{young-sp}, are described by Young tableaux with $N$ boxes
having no more than $n$ rows.
Let us introduce a partial order in the set of $\sun$ Young tableaux.

\begin{definition}
\label{def:ordering}
Let $\yn_1$ and $\yn_2$ be two Young tableaux with the \emph{same} number of boxes.
We set $\yn_1\succ\yn_2$ if $\yn_2$ may be obtained from
$\yn_1$ by the displacement of some of its boxes from the upper rows to the lower ones.
\end{definition}

\begin{example}
In the tableaux below the letters $a$ and $b$ mark the boxes, which change their positions.
We have for any $n\ge 3$:
\be
\label{movebox}
\newcommand{\mb}{\mbox{}}
{\Yvcentermath1
{\young(\mb\mb{a}{b},\mb)} \qquad \succ \qquad
{\young(\mb\mb{a},\mb{b})} \qquad \succ \qquad
{\young(\mb\mb,\mb{b},{a})} \;\; .
}
\ee
\end{example}

\begin{remark}
Note that not all Young tableaux can be compared in this way for higher ($n>2$) algebras.
For example, the two tableaux
$\yn_1={\tiny\Yvcentermath1 \yng(4,1,1)}\;$ and $\yn_2={\tiny\Yvcentermath1 \yng(3,3)}\;$
with six boxes each
%${\tiny\Yvcentermath1 \yng(2,1,1)}$ and ${\tiny\Yvcent ermath1 \yng(3,1)}$
are not related to each other by the partial order defined above.
\end{remark}

\begin{definition}
\label{def:yn-parity}
We'll say that the Young tableau $\yn$ has a certain parity (even or odd)
if the length parities of all rows in $\yn$ are equal
(are even or odd correspondingly).
\end{definition}

Now, we are ready to formulate the main result of this article.

\begin{theorem}
\label{tm:main}
Denote by $\en(\yn)$ the relative ground state energy on
the $\yn$ multiplet sector $\sptot_\yn$ of $N$-site chain \eqref{H}.
Then for open chains:
\begin{enumerate}
\renewcommand{\labelenumi}{(\roman{enumi})}
\item $\en(\yn)$ increases with respect to the partial order for $\yn$.
In other words, if $\yn'\succ\yn$ then $\en(\yn')>\en(\yn)$.
\item The relative ground state energy levels are nondegenerate in the sense
that among all $\yn$ multiplets only one has the lowest value $\en(\yn)$.
\end{enumerate}
For periodic chains, the two aforementioned conditions are fulfilled if, in addition,
the Young tableau $\yn$ has a certain parity in the sense of Definition~\ref{def:yn-parity}.
\end{theorem}

The theorem allows to get information about the degeneracy and $\sun$ structure
of the total ground state of finite chain.

\begin{corollary}
\label{cor:main}
The space spanned by all ground states of $N$-site chain \eqref{H} forms:
\begin{enumerate}
\renewcommand{\labelenumi}{(\roman{enumi})}
\item one-dimensional $\sun$ singlet, if $N\equiv 0\pmod{n}$, both for periodic
      and open chains;
\item $n$-dimensional fundamental $\sun$ multiplet, if $N\equiv 1\pmod{n}$, for open chains;
\item $C_n^m$-dimensional $m$-th order
      antisymmetric $\sun$ multiplet,  if $N\equiv m\pmod{n},1<m\le n-1$ , for open chains.
\end{enumerate}
\end{corollary}

\begin{proof}
For $N$-site chain denote by $\yn_\text{gs}$ the Young tableau, whose all columns have $n$ boxes
besides the last one, which has $m\equiv N\pmod{n}$ boxes. From Definition~\ref{def:ordering}
of the partial order it is evident that $\yn_\text{gs}\prec\yn$ for any other Young tableau $\yn$.
Note that $\yn_\text{gs}$ corresponds to $C_n^m$-dimensional antisymmetric $\sun$ representation
(see Section~\ref{sec:sln}). If $m=0$ (one-dimensional $\sln$ singlet,
$\yn_\text{gs}$ is $n\times(N/n)$ rectangle) then $\yn_\text{gs}$ is either even or odd, because
its all rows have the same length ($=N/n$).
The application of Theorem~\ref{tm:main} completes the proof.
\end{proof}

\begin{remark}
In case of $\su{2}$-symmetry, \eqref{H} is equivalent to
$\sum_{l=1}^{D(N)}J_{l\,l+1} \spin_\ind{l}\spin_\ind{l+1}$, which
corresponds to well-known Heisenberg
antiferromagnetic chain. Here $\spin=(S^x,S^y,S^z)$ is the spin operator on a site.
The Hamiltonian is invariant on the subspace of constant value of the total spin operator
$\spin=\sum_{l=1}^{N}\spin_\ind{l}$.

As has been  mentioned in Introduction,
for the Heisenberg antiferromagnet on bipartite lattice with rather
general Hamiltonian  Lieb and Mattis \cite{LM62,Lieb89}
proved exactly that among all spin-$S$ multiplets only one exists with minimal
eigenvalue $\en(S)$.
Remember that the lattice is called bipartite
if it can be separated into two disjoined sublattices $\mathcal{A}$ and $\mathcal{B}$
in such a way that the interactions present between the spins
from different sublattices only. The open chain and the periodic
chain with even number of sites are the examples of bipartite lattice.
The ground state of this Hamiltonian has total spin
$S_\text{gs}=|S_\mathcal{A}-S_\mathcal{B}|$,
where $S_\mathcal{A}$ and $S_\mathcal{B}$  are the maximum possible spins of the
two subsystems.
In addition, the lowest energy level $\en(S)$
for each spin value $S\ge S_\text{gs}$ is ordered in a natural way, i.e.
if $S_1>S_2\ge S_\text{gs}$ then $\en(S_1)>\en(S_2)$.
It follows from their result, in particular, that the ground state
of spin-$1/2$ open or periodic chain with even number of spins ($S_\mathcal{A}=S_\mathcal{B}$)
is spin-singlet and unique. At the same time the ground state
of open chain with odd number of spins ($S_\mathcal{A}=S_\mathcal{B}\pm1/2$) is
spin-doublet $S_\text{gs}=1/2$ and twofold degenerate.
So it is clear that the condition $S\ge S_\text{gs}$ is fulfilled for $S=1/2$ chains.
Hence, Theorem~\ref{tm:main} and Corollary~\ref{cor:main} are in agreement
with the energy level ordering established before \cite{LM62,Lieb89}.
\end{remark}

\section{$\sln$ algebra and its representations}
\label{sec:sln}

\subsection{Definition and fundamental representation}
\label{sec:sln-1}
As soon as we are dealing in this article with complex representations, the
complex extension of the real Lie algebra $\sun$, which coincides with the algebra
$\sln$ of all traceless complex $n\times n$ matrices, should be used instead.

The algebra $\sln$ consists of the diagonal part $\cartan$, called Cartan subalgebra,
upper $\borel_+$ and down $\borel_-$ triangular matrices, called Borel subalgebras:
$\sln=\borel_-\oplus\cartan\oplus\borel_+$.
\begin{align}
\label{sln}
\cartan  =  \spann \{\,e_{ii}-e_{i+1\,i+1}\mid 1\le i\le n-1\},
\quad
\borel_+ = \spann \{\,e_{ij}\mid 1\le i<j\le n\},
\quad
\borel_- = \spann \{\,e_{ij}\mid 1\le j<i\le n\},
\end{align}
where $e_{ij}$ ($i,j=1,\dots,n$) are $n\times n$ matrices, acting on the complex
space $V_n$, with single non-zero entry
at $i$-th row and $j$-th column: $e_{ij}:=\bra{i}\ket{j}$.
In fact, the generators in \eqref{sln} are given in the {\em defining} or {\em fundamental}
representation $\rep^\fund$ of $\sln$ algebra.
In the field theory language, the basic states $\bra{1},\dots,\bra{n}$ are called often
the color states. The $i$-th color is assigned to the basic state $\bra{i}$.
In case of $\sll{2}$, the two colors correspond to spin-up and spin-down states.

Let $\eps_i$ be the basic of linear forms dual to $e_{ii}$:
$\eps_i(e_{jj})=\delta_{ij}$ ($i,j=1,\dots,n$). Then the sets of positive
$\Delta_+ \subset h^*$, negative $\Delta_-\subset h^*$ and all $\Delta\subset h^*$ roots are
\begin{align}
\label{fund2}
 \Delta_+=\{\alpha_{ij}\mid 1\le i<j\le n\}, \quad
 \Delta_-=\{\alpha_{ji}\mid 1\le i<j\le n\}, \quad
\Delta=\Delta_-\cup\Delta_+,
\quad \text{where} \quad
\alpha_{ij}=\eps_i-\eps_j.
\end{align}
For any $\alpha=\alpha_{ij}$, $i\ne j$ we use another notations: $e_{\alpha}:=e_{ij}$
and $h_\alpha:=e_{ii}-e_{jj}$. Thus, the positive roots $\alpha\in\Delta_+$
(negative roots $\alpha\in\Delta_-$) correspond to the upper
Borel subalgebra $e_\alpha\in\borel_+$ (lower Borel subalgebra $e_\alpha\in\borel_-$).
Note also, if $\alpha\in\Delta_+$ then $-\alpha\in\Delta_-$ and vise versa.
The roots $\pm\alpha_i$, where
$\alpha_i:=\alpha_{i\,i+1}$, ($i=1,\dots,n-1$), are called simple roots.
Any $\alpha\in\Delta_\pm$ can be expressed in terms of simple roots with
integer coefficients as $\alpha=\pm\sum_{i=1}^{n-1} n_i\alpha_i$, where
$n_i\in\mathbb{Z}$, $n_i\ge0$.

The basis, generated by $h_{\alpha_i},e_\alpha$, $\alpha\in\Delta$, $i=1,\dots,n-1$
is known as the Cartan-Weyl basis.
The commutations in this basis have the following simple form:
\begin{align}
\begin{array}{l}
\com{h}{e_\alpha}=\alpha(h)e_\alpha, \quad \forall\,h\in \cartan,\\
\\
\com{h_1}{h_2}=0, \quad \forall \,h_1,h_2\in\cartan,
\end{array}
\qquad
\com{e_\alpha}{e_\beta}=
\begin{cases}
h_\alpha,         & \text{if $\beta=-\alpha$ and $\alpha\in\Delta_+$}, \\
0,                & \text{else if $\alpha+\beta\notin\Delta$}, \\
e_{\alpha+\beta}, & \text{if $\alpha+\beta\in\Delta$}
\end{cases}
\end{align}

The basic colors $\bra{i}\ i=1,\dots,n$ are {\it weight states}
with the {\it weights} $\eps_i$ each. This means that
$\rep^\fund(h)\bra{i}=\eps_i(h)\bra{i}$ for all $h\in\cartan$. The
{\it highest weight state}, which is annihilated by all elements
$\rep^\fund(e_\alpha)$, where $\alpha\in\Delta_+$, is the state
$\bra{1}$ and the highest weight is $\eps_1$. Other basic states
could be obtained from $\bra{1}$ by successive application of lowering
elements: $V_n=\rep^\fund(\borel_-)\bra{1}$. Note that the elements
$e_\alpha\in\borel_+$ ($e_\alpha\in\borel_-$) increases
(decreases) the color.

\begin{remark}
Sometimes, in the representation theory of Lie algebras all $m$-th order
($1\le m\le n-1$) antisymmetric $\sln$ representations are called fundamental.
In this article, however, we use more convenient notations for our purposes.
The fundamental representation, as we defined here, is unique, corresponds to
$m=1$ case and is given by \eqref{sln}.
\end{remark}

\subsection{Tensor product, Young tableaux and irreducible representations}
\label{sec:sln-2}

It is well known that the tensor products of different number of
fundamental representations contain all irreducible
$\sln$ multiplets. As it was mentioned above, these multiplets are
described by Young tableaux. Let us describe this procedure in
detail.

For the convenience, the two kinds of parametrization for $\sln$ Young tableau $\yn$
are used in this article: parametrization by row lengths and by column lengths.

\begin{notation}
Let $\{N_1,\dots,N_\nu\}$ be a partition of $N$, i.e. $N_1+\dots+N_\nu=N$, of length $\nu\le n$
given in nonascending order $N_1\ge\dots\ge N_\nu>0$.
Denote by $\yn[N_1,\dots,N_\nu]$ (in square brackets) the Young tableau with $N$ boxes
and $\nu$ rows,  $i$-th row containing $N_i$ boxes. Similarly, denote
by $\yn(N_1,\dots,N_\nu)$ (in standard brackets) the Young tableau with $N$ boxes
and $\nu$ columns,  $i$-th column containing $N_i$ boxes.
\end{notation}

Remember that on the product space $\sptot$ \eqref{young-sp},
the $\sln$ elements act in the standard way:
\be
\label{tensor}
\rep^{N}(x)=\sum_{i=0}^{N-1}x_\ind{i},\quad \text{where}\quad
x_\ind{i}=\rep^\fund_\ind{i}(x)=\id\otimes\dots\otimes\id\otimes \underbrace{\rep^\fund(x)}_i
\otimes\id\otimes\dots\otimes\id
\quad \text{and} \quad x\in\sln
\ee
The action $\rep^{N}$ is invariant with respect to the
permutation algebra. The irreducible representations,
contained in $\sptot$, are obtained after the appropriate
symmetrization-antisymmetrization procedure described below.
Distribute the $N$ indexes, corresponding to different $V_n$ in the tensor
product $\sptot$, among the boxes of $\yn[N_1,\dots,N_\nu]$.
We obtain an \emph{indexed} Young tableau, each box of which is
assigned to some site.
Then carry out the symmetrization over all indexes in each row
separately with the subsequent antisymmetrization over the
columns. The resulting multiplet will be irreducible. The
multiplets, constructed in this way from the Young tableaux of
different shapes are orthogonal to each other and belong to different
equivalence classes,
while the multiplets obtained from the same Young tableau $\yn$, are equivalent
and form the sector $\sptot_\yn$ as was already mentioned in Section \ref{sec:main}.
The different subspaces $\sptot_\yn$ in the decomposition \eqref{young-sp}
are orthogonal.

\begin{example}
Let $N=3$ and consider the $\yn[2,1]={\tiny\yc\yng(2,1)}$ tableau. The representation,
constructed from the index distribution
\raisebox{-0.4\bs}{\hoogte=0.8\bs\breedte=\bs\dikte=0.1pt%
\begin{Young}
 ${\scriptstyle i_1}$ & ${\scriptstyle i_3}$ \cr ${\scriptstyle i_2}$ \cr
\end{Young}
}
corresponds to the symmetrization of first and third
indexes and the subsequent antisymmetrization of the first and second ones.
\end{example}

\begin{remark}
All columns in Young tableau have to contain no more than $n$ boxes
to ensure the antisymmetrization over the column indexes does not vanish.
This fact explains the condition $\nu\le n$ we used above for $\yn[N_1,\dots,N_\nu]$.
Note that the $n$-length columns can be eliminated from $\yn$, because they produce
the trivial one-dimensional $\sun$ singlet. This gives rise to lower order tensor
representation. However, as was mentioned in Section~\ref{sec:main},
we don't cut the columns and consider the $N$-box Young tableaux only, because
in this article  we consider the multiplets, which appear in the decomposition
\eqref{young-sp} of the tensor product of $N$ fundamental representations.
\end{remark}

Now we will describe in detail the irreducible representation $\rep^\yn$,
corresponding to Young tableau $\yn$. Assigning to every box in $\yn$
a basic state $\bra{k}$ ($k=1,\dots,n$) from the corresponding site
and performing the symmetrization-antisymmetrization procedure as it was
described above, we will obtain some state from the multiplet $\rep^\yn$.
Thus, to any distribution of $n$ colors among the Young tableau boxes,
i.e. to any \emph{colored} Young tableau, a state (possibly, vanishing) of
the multiplet $\rep^\yn$ corresponds.
Among them one can choose the following basic states.

\begin{definition}
\label{def:young-basis}
The {\em standard basic states} in $\rep^\yn$ are those, whose
color values inside the boxes of $\yn$ don't decrease in left-to-right direction and
increase in up-to-down direction.
\end{definition}

Other states either are expressed in terms of the basic ones or are vanished.

\begin{example}
\label{ex:bvec}
Consider $\sll{3}$ algebra acting on $(N=3)$-rd order tensor space $\sptotN{3}$.
The representation $\rep^{\yn[2,1]}$ is eight-dimensional and the following
states form the standard basis there:
\be
\label{bvec}
{\yc \young(11,2) \qquad \young(11,3) \qquad \young(12,2)\qquad \young(12,3)\qquad \young(13,2)
\qquad \young(13,3) \qquad \young(22,3) \qquad \young(23,3)  }
\ee
\end{example}

\begin{example}
The fundamental representation $\rep^\fund$ \eqref{sln} is
described by the single box tableau $\yn[1]=\yn(1)=\Box$ and has $n$
states ${\yc\young(k)}=\bra{k}$, where $k=1,\dots,n$. The Young
tableau $\yn(m)$ consisting of one column
describes the $m$-th order antisymmetric representation with dimensions
$C_{n}^{m}=n!/(m!(n-m)!)$ provided that $2\le m\le n-1$.
\end{example}

The above mentioned standard basic states constitute the weight space
states of $\rep^\yn$, i.e.
they are diagonal with respect to the Cartan subalgebra $\cartan$ of $\sln$.
Namely, any state $\Psi^\lambda_\yn$ from the standard basis obeys the equation
$\rep^\yn(h)\Psi^\lambda_\yn=\lambda(h)\Psi^\lambda_\yn$, where $h\in \cartan$
and the weight $\lambda$ is expressed in terms of the
linear forms $\eps_i$ defined in Section~\ref{sec:sln-1} as
\be
\label{weight}
\lambda=\lambda_{\widetilde{N}_1\dots\widetilde{N}_n}=
\sum_{i=1}^n \widetilde{N}_i\,\eps_i,
\ee
where $\widetilde{N}_i$ is the number of boxes in $\yn$ filled with $i$-th
color.
Among the basic states,
there is the unique highest weight state $\Psi_\yn=\Psi^{\lambda_\yn}_\yn$, which is
annihilated by all rising elements $e_\alpha\in\borel_+$.
The corresponding highest weight
$\lambda_{\yn}$ has the coefficients $\widetilde{N}_i$
equal to row lengths $N_i$ of $\yn=\yn[N_1,\dots,N_\nu]$ for $i\le \nu$ and
are zero for $i>\nu$:
$\lambda_{\yn}=\sum_{i=1}^\nu N_i\eps_i$.
This known result is used in the article so below its complete proof
is given.

\begin{proposition*}[The structure of the highest weight state]
The highest weight state $\Psi_\yn$ corresponds to such
distribution of colors in $\yn$ that all boxes in $i$-th row have
the same color $i$. (See Fig.~\ref{fig:hvec}(a))
\end{proposition*}

\begin{figure}
$$
\text{\raisebox{0.5\bs}{\large (a)}}
\quad
%\gs_{\yn[1]}={\yc\young(1)}=\bra{1} \qquad
\Psi_{\yn[3]}={\yc\young(111)} \qquad
\Psi_{\yn[1,1,1,1]}={\yc\young(1,2,3,4)} \qquad
\Psi_{\yn[3,2,1]}={\yc \young(111,22,3)}
\qquad
\qquad
\text{\raisebox{0.5\bs}{\large (b)}}
\quad
\Psi_{\yn[3,2,1]}=%
\text{%
\raisebox{-1.5\bs}%
{\begin{Young}
${ 1_1}$ & $1_4$ & ${1_6}$ \cr
${ 2_2}$ & $2_5$ \cr
${ 3_3}$ \cr
\end{Young}}
}
$$
\caption{{\sf (a)}~The highest weight states $\Psi_\yn$ for different type Young tableaux.
The $i$-th row should be filled by $i$-th color.
{\sf (b)}~The highest weight state constructed in \eqref{eq:hvec} for $\rep^{\yn[3,2,1]}$
($N=6$, $n\ge 3$). The site's index, assigned to each box, is given in the subscript.%
}
\label{fig:hvec}
\end{figure}

\begin{proof}
Note that the symmetrization over the rows of
$\yn=\yn[N_1,\dots,N_{M_1}]=\yn(M_1,\dots,M_{N_1})$ is trivial for
the states of this type. So, the antisymmetrization for each
column remains to be performed only. Without the loss of
generality one can suppose that the first $N_1$ sites of $\sptot$
are assigned to the first column of $\yn$ in ascending order in
downward direction, next $N_2$ sites to the second one, etc. See
Fig.~\ref{fig:hvec}(b) for the example. Then for the highest
weight state of the constricted multiplet we have: \be
\label{eq:hvec} \Psi_{\yn(M_1,\dots,M_{N_1})}=\sqrt{M_1!\dots
M_{N_1}!}\;\bigotimes_{i=1}^{N_1}\;
        \asym_{M_i}\bra{1\,2\ldots M_i-1\, M_i}.
\ee
Here $\asym_{M}$ is the antisymmetrization operator
\be
\label{Q-m}
\asym_{M}\bra{k_1\,k_2\ldots k_{M-1}\, k_M}
=\frac{1}{M!}\sum_{\{p_i\}\in \gperm(M)}\epsilon_{p_1p_2\ldots p_{M-1} p_{M}}
\bra{k_{p_1}\,k_{p_2}\ldots k_{p_{M-1}}\, k_{p_{M}}},
\ee
where $\epsilon_{p_1\dots p_M}$
is the standard $M$-th order absolutely antisymmetric tensor and by $\gperm(M)$ the group of
the permutations of successive numbers $\{1,\dots,M\}$ is denoted.
Thus, \eqref{eq:hvec} decouples on tensor
product of $N_1$ parts, each consisting of anti-symmetrization of
the tensor product $\bra{1\,2\ldots M_i-1\, M_i}$. The rising
elements $\rep^N(e_\alpha)$, $\alpha\in\Delta_+$, acting on
each multiplier factor in the right hand side of \eqref{eq:hvec}
vanish (see \eqref{tensor}):
\be
\label{ehv}
     \sum_{m=1}^{M_i}\rep^\fund_\ind{m}(e_\alpha)
     \asym_{M_i}
     \bra{1\,2\ldots M_i-1\, M_i}
     =
     \asym_{M_i}
     \sum_{m=1}^{M_i}\rep^\fund_\ind{m}(e_\alpha)
     \bra{1\,2\ldots M_i-1\, M_i}
     \,=\,0
\ee
Indeed,  $\rep^\fund(e_\alpha)\bra{k}$ equals either to zero or
to some $\bra{k'}$ with $k'<k$. In the second case, the corresponding term
of the sum in \eqref{ehv} vanishes too due to the antisymmetrization over two
equal states $\bra{k'}$ performed in \eqref{ehv}.
\end{proof}
%

%\begin{remark}
%As was mentioned above, the attachments of $n$-length column to $\yn$ doesn't change the
%representation's equivalence class. This corresponds to the addition
%to \eqref{weight} of the element $\sum_{i=1}^n\eps_i$, which doesn't belong to $\cartan^*$.
%The form $\lambda'_{}=\lambda-1/N\sum_{i=1}^N\eps_i$
%is the unique element of $\cartan^*$, which describes the equivalence class of $\rep_{\yn[N_1,\dots,N_\nu]}$.
%\end{remark}
%
%

\section{Relative ground states on weight subspaces}
\label{sec:relative}

%\subsection{Preliminary notations}

For our purposes it is convenient to use another form of the Hamiltonian \eqref{H}.
For the chain with fundamental representation considered in this article,
the nearest-neighbor interactions in \eqref{H} are expressed in terms of
the permutations:
\be
  \label{HP}
  \hamt=\sum_{l=1}^{D(N)}J_{l\,l+1} \perm_\ind{l\,l+1}=\ham+\frac1n\sum_{l=1}^{D(N)}J_{l\,l+1},
\ee
where $\perm_\ind{ij}$ permutes the colors of $i$-th and $j$-th
sites resting other sites unchanged:
$$
\perm_\ind{ij}\;\bras{\dots \underset{i}{k_i}\dots \underset{j}{k_j}\dots}
=
\bras{\dots \underset{i}{k_j}\dots \underset{j}{k_i}\dots}
$$
and $D(N)$ is defined in \eqref{H}.
Thus $\hamt$ differs from $\ham$ by a non-essential scalar term.

For any sequence $N_i\ge0$ obeying $N_1+\dots+N_n=N$, we
define by $\sptot_{N_1\dots N_n}$ the subspace in $\sptot$ spanned by
states $\bra{k_1\dots k_N}$, which contain $N_1$ colors of type $1$,
$N_2$ colors of type $2$, etc., i.e.
$\{\,\#\,i\; |\; k_i=k,\,1\le i\le N\,\}=N_k$.
In other words using \eqref{tensor}:
\be
\label{subsp}
\sptot_{N_1\dots N_n} := \spann\left\{\ \bra{k_1\dots k_N}\
\right| \ \left.
\rep^N(e_{ii}) \bra{k_1\dots k_N} = N_i \bra{k_1\dots k_N},\ 1\le i\le n \right\}.
\ee
This subspace is diagonal with respect to the Cartan subalgebra:
$\rep^N(h)|_{\sptot_{N_1\dots N_n}}=\lambda_{N_1\dots N_n}(h)\cdot\id$ for any
$h\in\cartan$. It is called a \emph{weight} subspace corresponding to weight
$\lambda_{N_1\dots N_n}:= \sum_{i=1}^n N_i\eps_i$.
Given one representative $\bra{k_1\dots k_N}\in\sptot_{N_1\dots N_n}$ any
other state  $\bra{k'_1\dots k'_N}\in\sptot_{N_1\dots N_n}$
can be obtained from it by some permutation $\{p_i\}\in\gperm(N)$ as
$\bra{k'_1\dots k'_N}=\bra{k_{p_1}\dots k_{p_N}}$.

The defined subspaces constitute together the total space of the states:
\be
\label{subsp-sum}
%\sptot = \bigoplus_{\substack{N_i\ge 0, \\ N_1+\dots+N_n=N }}\sptot_{N_1\dots N_n}.
\sptot = \bigoplus_{N_i\ge 0}^{N_1+\dots+N_n=N }\sptot_{N_1\dots N_n}.
\ee
Due to the symmetry with respect to the diagonal subalgebra $\cartan$,
the Hamiltonian \eqref{HP} is \emph{invariant} on
the subspaces $\sptot_{N_1\dots N_n}$. Hence, its restriction
$\hamt|_{\sptot_{N_1\dots N_n}}$ on $\sptot_{N_1\dots N_n}$ is well defined.
We call below the minimal energy states of $\hamt|_{\sptot_{N_1\dots N_n}}$ as
the \emph{relative} ground states in contrast to the total
ground state of $\hamt$ on the entire space $\sptot$.

\subsection{Uniqueness condition for the relative ground states}

Before studding the properties of minimal energy states in the subspaces
$\sptot_{N_1\dots N_n}$ we provide the following definition for the parity
of the sequence $\{k_1,\dots, k_N\}$.

\begin{definition}
\label{def:parity}
We say that for $i<j$, $i$-th and $j$-th elements in sequence $\{k_1,\dots, k_N\}$
are \emph{disordered} if $k_i>k_j$, otherwise ($k_i\le k_j$) they are \emph{ordered}.
Define the \emph{parity} $\sigma_{k_1\dots k_N}=\pm 1$ of the sequence $\{k_1,\dots, k_N\}$
as the parity of the number of all disordered pairs in it:
$
\sigma_{k_1\dots k_N} = (-1)^{ \{\, \#(i<j) \; | \; k_i>k_j\, \} }
$.
\end{definition}

Now we formulate the main result of the current section.

\begin{lemma}
\label{lm:subspace}
The Hamiltonian \eqref{HP} has unique ground state on
each subspace $\sptot_{N_1\dots N_n}$ in case of an open chain.
In case of a periodic chain, the ground state on $\sptot_{N_1\dots N_n}$
is unique, if all nonzero color multiplicities $N_1,\dots,N_n$
have the same parity (even or odd):
$N_i\equiv N_j\pmod{2}$ if $N_iN_j>0$.
\end{lemma}

\begin{proof}
We'll prove the lemma in three steps. First, we will find the basis, in which all
nonzero off-diagonal matrix elements of $\hamt$ are negative. Then we'll show that the
corresponding matrix is connected (see Definition~\ref{def:connectivity} of
connected matrix in Appendix.
In the last step, by applying the Perron-Frobenius theorem (see Appendix),
we'll conclude that the ground state is unique.

The parity introduced in Definition~\ref{def:parity} above has the following nice property:
for open chains, it changes the sign under the permutation of neighboring colors
provided that they differ from each other:
\be
\label{sigma}
\sigma_{k_1\dots k_ik_{i+1}\dots k_N}=
  (-1)^{1-\delta_{k_ik_{i+1}}}\sigma_{k_1\dots k_{i+1}k_i\dots k_N}.
\ee
In order to make sure of this, note that $(k_i,k_{i+1})$ is the only pair
of colors, which changes the ordering under the permutation $k_i\leftrightarrow k_{i+1}$.
Now it is easy to see that the basis we are looking for is
\be
\label{basis}
\brav{k_1\dots k_N}=\sigma_{k_1\dots k_N} \bra{k_1\dots k_N},
\quad \text{where}  \quad
\bra{k_1\dots k_N}\in \sptot_{N_1\dots N_n}
\ee
Indeed, since $\hamt$ \eqref{HP} is a sum of nearest-neighbor permutations
with positive coefficients ($J_{l\,l+1}>0$), its nonzero off-diagonal elements in the basis
\eqref{basis} have to be negative.

It is evident that every two basic elements of this type in $\sptot_{N_1\dots N_n}$
could be connected by successive permutations of neighboring colors,
which proves the connectivity of $\hamt\,|_{\sptot_{N_1\dots N_n}}$ in this basis.
The application of the well known Perron-Frobenius theorem,
which is given with the proof in Appendix,
finishes the proof for the chains with the open boundary
conditions.

For the rings, however, $\hamt$ can have positive off-diagonal elements
in the basis \eqref{basis} due to the last, $N$-th term in \eqref{HP}.
Indeed, the assertion that the parity changes the sign under the permutation
of the first and the last colors if they differ from each other is not true, in general.
It is valid, however, provided that the nonzero color multiplicities have the same parity.
Indeed, suppose that $k_1\ne k_N$ and exchange $k_1$ and $k_N$
in $\sigma_{k_1\dots k_N}$ using the successive nearest-neighbor permutations.
While permuting with $i$-th color, which differs from $k_1$ and $k_N$
($k_i\ne k_1$, $k_i\ne k_N$), both $k_1$ and $k_N$ produce sign factors
$-1$, which are cancelled out together. Hence, it remains to take into account
the permutations with the colors equal to either $k_1$ or $k_N$:
\be
\label{k1kN}
\sigma_{k_1k_2\dots k_{N-1} k_N}
=(-1)^{N_{k_1}+N_{k_N}-1}\sigma_{k_Nk_2\dots k_{N-1} k_1}, \quad
\text{if} \quad k_1\ne k_N
\ee
Thus, the minus
sign occurs, if and only if all nonzero $N_i$ have the same parity.
The proof for the rings is completed also.
\end{proof}

\begin{remark}
\label{rem:gs}
Let $\gs_{N_1\dots N_n}\in \sptot_{N_1\dots N_n}$ be the relative
ground state. During the proof of Lemma~\ref{lm:subspace}
the basis \eqref{basis} in this subspace has been found
where all nonzero off-diagonal elements of $\hamt$ are negative.
The Perron-Frobenius theorem asserts also (see Appendix)
that all coefficients of $\gs_{N_1\dots N_n}$ in this basis should
be \emph{strongly} positive:
\be
\label{gs}
\gs_{N_1\dots N_n}=
%\sum_{\{\#i |k_i=k\}=N_k,\, 1\le k\le n}
\sum_{1\le k_1,\dots,k_N\le n}^{\bra{k_1\dots k_N}\in\sptot_{N_1\dots N_n}}
\omega_{k_1\dots k_N} \brav{k_1\dots k_N},
\qquad \text{where \quad $\omega_{k_1\dots k_N}>0$}
\ee
This property of the relative ground state will be used below.
\end{remark}

\begin{remark}
In $\su{2}$ case, there are two colors and $\sptot_{N_1N_2}$ is formed by the
states with $N_1$ spin-up and $N_2$ spin-down sites, i.e. corresponds to
$S^z=(N_1-N_2)/2$ subspace.
If the ring contains even number of sites ($N=2N'$),
then both $N_1$ and $N_2$ have the same parity and all spin $S^z$ subspaces
have unique ground states. This is in accordance with the known results about
$\sll{2}$ symmetric chains on a bipartite lattice.
If, however, the ring contains odd number of sites ($N=2N'+1$, the lattice is not
bipartite), then the parities of $N_1$ and $N_2$ differ. Recent numerical investigations
show that the ground states in this case are in fact degenerate
\cite{Schm00,Schm03}.
Hence, one could suppose, that the condition of Lemma~\ref{lm:subspace} for
the periodic chains is strong enough.
\end{remark}

\begin{remark}
In fact Lemma~\ref{lm:subspace} possesses generalization to bipartite
lattices for $\su{2}$ symmetric models \emph{only}. Indeed, let
$\mathcal{A}$ and $\mathcal{B}$ are two disjoined sublattices of bipartite
lattice $\mathcal{L}$.
Enumerate the sites of $\mathcal{A}$ by successive odd numbers
($i_\mathcal{A}=1,3,5,\dots$)
and the sites of $\mathcal{B}$ by successive even numbers
($i_\mathcal{B}=2,4,6,\dots$). It is easy to verify that the spin exchange
between $i$-th and $j$-th sites affects on the the parity as
$
\sigma_{\dots k_i\dots k_j \dots}
=(-1)^{i-j} \sigma_{\dots k_j\dots k_i \dots}
$
provided that $k_i\ne k_j$. Note that this property is valid for $n=2$ only,
where $k_i=1,2$. Since the interaction presents between the even
and odd sites only, we conclude that all off-diagonal matrix elements of the
$\su{2}$ symmetric Hamiltonian in the basis \eqref{basis} on bipartite lattice
are negative.
\end{remark}

\section{Relative ground states on the multiplet subspaces $\sptot_\yn$}
\label{sec:ordering}

\subsection{Multiplets containing relative ground states $\gs_{N_1\dots N_n}$ }

In the previous section, using the symmetry of $\hamt$ with respect to the diagonal
subalgebra $\cartan$ of $\sln$,
the ground state degeneracy of $\hamt$ on the weight subspaces $\sptot_{N_1\dots N_n}$
have been studied.
However, as was already mentioned, the Hamiltonian possesses the symmetry with respect
to entire $\sln$ algebra and has block-diagonal form with respect to the decomposition
\eqref{young-sp} on non-equivalent multiplet sectors $\sptot_\yn$.
To which multiplet sector does belong the relative ground state $\gs_{N_1\dots N_n}$?
The following lemma adverts to this question.
%First consider the case when the sequence of color multiplicities is given in
%nonascending order.

\begin{lemma}
\label{lm:multiplet}
For the open chain, the relative ground state $\gs_{N_1\dots N_n}$ belongs to the
$\yn$ multiplet sector $\sptot_{\yn}$, where
$\yn=\yn[N_{p_1},\dots,N_{p_\nu}]$, the permutation
$\{p_i\}\in\gperm(n)$ is chosen to satisfy $N_{p_1}\ge\dots\ge N_{p_n}$ and
$\nu\le n$ is the number of nonzero $N_i$, i.e. $N_{p_\nu}>0$ and $N_{p_{\nu+1}}=0$.
For the periodic chain, the condition
$\gs_{N_1\dots N_n}\in\sptot_{\yn}$ is fulfilled if all $\nu$
nonzero color multiplicities have the same parity.
\end{lemma}

\begin{proof}
According to Lemma~\ref{lm:subspace} and
under the given conditions, the relative ground state $\gs_{N_1\dots N_n}$ is
nondegenerate and all its coefficients in the basis
\eqref{basis} are positive as it was mentioned in Remark \ref{rem:gs}.

First, we'll construct a state $\Psi^{\{p_i\}}_\yn$, belonging to the sector
$\sptot_\yn$, which has nonzero overlap with the relative ground state
$\gs_{N_1\dots N_n}$.
Let $M_i$, $1\le i\le m=N_{p_1}$ is the length of the $i$-th column in
$\yn$. Using the notation from Section \ref{sec:sln-2} we can express the same
Young tableau as $\yn=\yn(M_1,\dots,M_m)$.
Attach to the box, which is located on $i$-th row and $j$-th column of
$\yn$ the ($i+\sum_{j'=1}^{j-1}M_{j'}$)-th site of the chain.
In other words, the site indexes increase successively along the columns.
Assign to each box of $i$-th row the \emph{same} color state $\bra{p_i}$.
Figure~\ref{fig:tildev0} describes this procedure on a concrete
example.
The state $\Psi^{\{p_i\}}_\yn$, obtained in this way,
is the tensor product of $m$ antisymmetrized parts corresponding to columns in
left-to-right direction:
\be
\label{overlap}
\Psi^{\{p_i\}}_\yn=
\sqrt{M_1!\dots M_m!}\;
\asym_{M_1}\bra{p_1\dots p_{M_1}}
\otimes
\asym_{M_2}\bra{p_1\dots p_{M_2}}
\otimes \ldots \otimes
\asym_{M_m}\bra{p_1\dots p_{M_m}}
\ee
Note that for the trivial permutation $p_i=i$, $1\le i\le n$, the state
above
$\Psi^{\{p_i\}}_\yn$ coincides with the expression of the highest weight state
of the current representation, given in \eqref{eq:hvec}:
$\Psi^{\{i\}}_\yn=\Psi_\yn$.
Using \eqref{Q-m}, \eqref{sigma} and \eqref{basis}, the state defined
above can be presented as follows:
%$$
%\asym_{M_j}\bra{p_1\,p_2\ldots p_{M_j-1}\, p_{M_j}}
%%=\sigma_{p_1\dots p_{M_j}}\asym_{M_j}\brav{p_1\,p_2\ldots p_{M_j-1}\, p_{M_j}}
%=\frac{\sigma_{p_1\dots p_{M_j}}}{M_j!}\sum_{\{q_i\}\in \gperm(M_j)}
%\brav{p_{q_1}\,p_{q_2}\ldots p_{q_{M_j-1}}\, p_{q_{M_j}}}
%$$
\begin{align}
\label{overlap-2}
\begin{split}
\Psi^{\{p_i\}}_\yn&=\frac{\sigma_{p_1\dots p_N}}{\sqrt{M_1!\dots M_m!}}
\sum_{\{q^{(j)}_i\}\in\gperm(M_j)}^{1\le j\le m}
\brav{p_{q^{(1)}_1}\dots p_{q^{(1)}_{M_1}}\;
p_{q^{(2)}_1}\dots p_{q^{(2)}_{M_2}}\;
\ldots \ldots\;p_{q^{(m)}_1}\dots p_{q^{(m)}_{M_m}}}
\\
&=
\frac{\sigma_{p_1\dots p_N}}{\sqrt{M_1!\dots M_m!}}
\sum_{\{q_i\}\in\gperm(N|M_1,\dots, M_m)}
\brav{p_{q_1}\ldots p_{q_N}},
\end{split}
\end{align}
where by $\gperm(N|M_1,\dots, M_m)$ the following subgroup of whole permutation group $\gperm(N)$
is denoted. Divide $N$ successive numbers on $m$ blocks, $j$-th
block containing $M_j$ numbers. The numbers are permuted resting each inside of its block.
By the construction we have: $\gperm(N|M_1,\dots, M_m)=\bigotimes_{j=1}^m\gperm(M_j)$.
Note that in the derivation of \eqref{overlap-2} we have used the fact that all numbers in each set
$\{p_1,\dots,p_{M_j}\}$, $1\le j\le m$, are different.
%$$
%\Psi^{\{p_i\}}_\yn=\mathcal{N}\;
%\sum_{\{q^{(j)}_i\}\in\gperm(M_j)}^{1\le j\le m}
%\brav{p_{q^{(1)}_1}\dots p_{q^{(1)}_{M_1}}\;
%p_{q^{(2)}_1}\dots p_{q^{(2)}_{M_2}}\;
%\ldots \ldots\;p_{q^{(m)}_1}\dots p_{q^{(m)}_{M_m}}}
%$$
%$$
%\Psi^{\{p_i\}}_\yn=\mathcal{N}\;
%\sum_{\{q^{j}_i\}\in\gperm(M_j)}^{1\le j\le m}
%\brav{p_{q^{1}_1}\dots p_{q^{1}_{M_1}}\;
%p_{q^{2}_1}\dots p_{q^{2}_{M_2}}\;
%\ldots \ldots\;p_{q^{m}_1}\dots p_{q^{m}_{M_m}}}
%$$
%$$
%\mathcal{N}=\prod_{j=1}^m \frac{\sigma_{p_1\dots p_{M_j}}}{\sqrt{M_j!}}
%    \prod_{1\le i<j\le m}\sigma(p_1\dots p_{M_i}|p_1\dots p_{M_j})
%$$

We conclude from \eqref{overlap-2} that the nonzero coefficients of the state $\Psi^{\{p_i\}}_\yn$
expressed in terms of the basis \eqref{basis} have the \emph{same} sign equal to $\sigma_{p_1\dots p_N}$.
According to Remark \ref{rem:gs} all coefficients of the relative ground state
in the same basis are positive.
Hence the scalar product of both states should be nonzero:
$\left(\Psi^{\{p_i\}}_\yn,\gs_{N_1\dots N_n}\right)\ne 0$.

Thus, we have constructed a state
$\Psi^{\{p_i\}}_\yn \in \sptot_\yn$,
which has an overlap with the relative ground state
$\gs_{N_1\dots N_n}$.
Suppose now that $\gs_{N_1\dots N_n}$ doesn't belong to $\yn$ multiplet sector
$\sptot_\yn$. This means that it contains a part
belonging to the other multiplet sectors, i.e. can be decomposed as:
$\gs_{N_1\dots N_n}=\mu \Psi^{\{p_i\}}_\yn + \Psi_\perp$,
where $\mu\ne0$ and $\Psi_\perp \perp \sptot_{\yn}$.
Since $\hamt$ has block-diagonal form with respect to the decomposition
\eqref{young-sp}, both $\Psi^{\{p_i\}}_\yn$ and
$\Psi_\perp$ should also be an eigenstates with the \emph{same} eigenvalue
as $\gs_{N_1\dots N_n}$. Obviously all these states belong to the weight
subspace $\sptot_{N_1\dots N_n}$.
This contradicts with the conditions of the uniqueness of the relative ground state
in $\sptot_{N_1\dots N_n}$ already proved in Lemma~\ref{lm:subspace}.
Thus, we conclude that $\gs_{N_1\dots N_n}\in\sptot_\yn$. This completes the proof.

\begin{figure}
%\hoogte=0.8\bs\breedte=\bs\dikte=0.1pt%
$$
%\yn[3,2,1]={\yc\yng(3,2,1)}
\Psi^{\{2,3,1\}}_{\yn[3,2,1]}=%
\text{%
\raisebox{-1.5\bs}%
{\begin{Young}
${ 2_1}$ & $2_4$ & ${2_6}$ \cr
${ 3_2}$ & $3_5$ \cr
${ 1_3}$ \cr
\end{Young}}
}
\qquad
\Psi^{\{1,3,2\}}_{\yn[3,2,1]}=%
\text{%
\raisebox{-1.5\bs}%
{\begin{Young}
${ 1_1}$ & $1_4$ & ${1_6}$ \cr
${ 3_2}$ & $3_5$ \cr
${ 2_3}$ \cr
\end{Young}}
}
\qquad
\Psi^{\{2,1,3\}}_{\yn[3,2,1]}=%
\text{%
\raisebox{-1.5\bs}%
{\begin{Young}
${ 2_1}$ & $2_4$ & ${2_6}$ \cr
${ 1_2}$ & $1_5$ \cr
${ 3_3}$ \cr
\end{Young}}
}
\qquad
\Psi^{\{3,1,2\}}_{\yn[3,2,1]}=%
\text{%
\raisebox{-1.5\bs}%
{\begin{Young}
${ 3_1}$ & $3_4$ & ${3_6}$ \cr
${ 1_2}$ & $1_5$ \cr
${ 2_3}$ \cr
\end{Young}}
}
$$
\caption{ The construction of states $\Psi^{\{p_1,p_2,p_3\}}_{\yn[3,2,1]}$ \eqref{overlap}
for different permutations $\{p_1,p_2,p_3\}\in\gperm(3)$,
each having nonzero overlap with the corresponding relative
ground state $\gs_{p_1p_2p_3}$ \eqref{gs},
in case of $\su{3}$ symmetric six-site chain.
The multiplet, to which all they belong, is characterized by the
Young tableau $\yn[3,2,1]$.
The site indexes, given as a subscripts with respect to the colors,
grow successively along the columns.}
\label{fig:tildev0}
\end{figure}
\end{proof}

According to Lemma~\ref{lm:multiplet}, for open chains
all relative ground states
in the weight subspaces $\sptot_{N_{p_1}\dots N_{p_n}}$ differing by
permutations $\{p_i\}\in\gperm(n)$ of color multiplicities belong to the same
multiplet sector $\sptot_\yn$, where $\yn=\yn[N_1,\dots,N_\nu]$ and
$N_1\ge\dots \ge N_n\ge0$. For rings the same is true for the weight
subspaces whose nonzero color multiplicities have the same parity.
Actually it appears that they all belong to one irreducible representation
$\rep^\yn$. Namely, the following proposition is valid.

\begin{proposition}
\label{prop:same}
Consider a partition of $N$ such that
$N_1\ge\dots \ge N_n\ge0$.
For open chain all relative ground states $\gs_{N_{p_1}\dots N_{p_n}}$, where
$\{p_i\}\in\gperm(n)$, belong to one multiplet
$V^\gs_\yn\subset\sptot_{\yn}$,
where $\yn=\yn[N_1,\dots,N_\nu]$ and $\nu\le n$ is the number of nonzero $N_i$,
i.e. $N_\nu>0$ and $N_{\nu+1}=0$.
The multiplet $V^\gs_\yn$ combines all lowest energy states in $\yn$ multiplet
sector $\sptot_\yn$.
For periodic chain the same result is true provided that the Young tableau $\yn$
has a certain parity in sense of Definition \ref{def:yn-parity}.
%$N_i=N_j\pmod{2}$ for all
%$1\le i,j\le \nu$.
\end{proposition}

\begin{proof}
Note that if the Young tableau $\yn[N_1,\dots,N_\nu]$ has a certain parity then
$N_i\equiv N_j\pmod{2}$ for all $1\le i,j\le \nu$, i.e. the condition of
Lemma~\ref{lm:multiplet} for rings is fulfilled.

According to Lemma~\ref{lm:multiplet}, the relative ground state
$\gs_{N_{p_1}\dots N_{p_n}}$ belongs to some irreducible
$\sln$ representation space $V_{\yn}^{\gs,\{p_i\}}\subset\sptot_\yn$.
It follows from the irreducibility of $V_{\yn}^{\gs,\{p_i\}}$ and $\sln$ symmetry
of $\hamt$ that
all the states of this representation are eigenstates of the Hamiltonian with the
\emph{same} eigenvalue $\en_{N_{p_1}\dots N_{p_n}}$:
$\hamt|_{V_{\yn}^{\gs,\{p_i\}}}=\en_{N_{p_1}\dots N_{p_n}}\cdot\id$.

From the other hand, the subspace $V_{\yn}^{\gs,\{p_i\}}$,
like any irreducible $\sln$ representation
described by Young tableau $\yn=\yn[N_1,\dots,N_\nu]$, contains the states
$\Psi^{\{q_i\}}_\yn$ \eqref{overlap} from the weight subspaces
$\sptot_{N_{q_1},\dots,N_{q_n}}$ for all permutations $\{q_i\}\in\gperm(n)$.
Suppose now that $\en_{N_{p_1}\dots N_{p_n}}<\en_{N_{1}\dots N_{n}}$.
This suggestion contradicts with the fact that $\gs_{N_{1}\dots N_{n}}$
is the relative ground state on the subspace $\sptot_{N_{1}\dots N_{n}}$.
Similarly, the inverse relation $\en_{N_{p_1}\dots N_{p_n}}>\en_{N_{1}\dots N_{n}}$
contradicts with the fact that $\gs_{N_{p_1}\dots N_{p_n}}$
is the relative ground state on
the subspace $\sptot_{N_{p_1}\dots N_{p_n}}$.

So, all eigenvalues have to be equal: $\en_{N_{p_1}\dots N_{p_n}}=\en_{N_{1}\dots N_{n}}$.
The uniqueness of the relative ground states $\gs_{N_{p_1}\dots N_{p_n}}$, proved in
Lemma~\ref{lm:subspace},
implies that all considered multiplets have to coincide: $V^\gs_\yn:\equiv V_{\yn}^{\gs,\{p_i\}}$.
Since any $\yn$ multiplet has a representative in weight subspace $\sptot_{N_{1}\dots N_{n}}$,
the space $V^\gs_\yn$ contains \emph{all} lowest
energy states belonging to $\yn$ multiplet sector.
In terms of the notation used in Theorem \ref{tm:main} we have: $\en_{N_{1}\dots N_{n}}=\en(\yn)$.
The proof is completed now.
\end{proof}

\begin{remark}
Obviously, the relative ground state $\gs_{N_1N_2\dots N_{n-1} N_n}$ is the highest weight state
of the multiplet $V^\gs_\yn$ while $\gs_{N_n N_{n-1}\dots N_{2} N_1}$, whose color
multiplicities are places in reverse, nondescending order, is its lowest weight state.
The later is a state annihilated by all lowering elements $e_\alpha\in\borel_-$ defined in
Section \ref{sec:sln}.
%For $\su{2}$ chain
\end{remark}

\subsection{The ordering of relative ground state energy levels $\en(\yn)$:
           proof of Theorem~\ref{tm:main}}

Proposition~\ref{prop:same}, in fact, classifies the relative ground states in
weight subspaces in terms of irreducible $\sln$  representations.
Namely, for any ordered partition $N_1\ge \dots \ge N_n\ge 0$ of $N$ the lowest
energy states in $\sptot_{N_{p_1}\dots N_{p_n}}$ for all
permutations $\{p_i\} \in \gperm(n)$ are combined in the same $\yn$ multiplet,
$\yn=\yn[N_1,\dots,N_n]$, with the energy eigenvalue $\en(\yn)$
corresponding to the minimal energy level among all $\yn$ multiplets.

However the following question remains open: how are
different relative ground state energy values $\en(\yn)$ related to each other?
In other words, could we compare the lowest energy values of two
different weight subspaces $\sptot_{N_1\dots N_n}$
and $\sptot_{N'_1\dots N'_n}$ if the sequences $\seq{N_1}{N_n}$
and $\seq{N'_1}{N'_n}$ don't come to each other by a
permutation? In this case, $\gs_{N_1\dots N_n}$ and
$\gs_{N'_1\dots N'_n}$ belong to different non-equivalent
representations $\rep^\yn$ and $\rep^{\yn'}$.
We advert to this problem in the current section.

The following proposition  indicates the conditions under which the lowest
energy levels $\en(\yn)$ and $\en(\yn')$ in different multiplet sectors
$\sptot_\yn$ and $\sptot_{\yn'}$ may be compared.

\begin{proposition}
\label{prop:ordering-1}
Consider two different $\sun$ Young tableaux, $\yn'=\yn[N'_1,\dots,N'_{\nu'}]$ being arbitrary
and \\
$\yn=\yn[N_1,\dots,N_\nu]$ being either even or odd in case of the ring,
where $\nu,\nu'\le n$.
Suppose that the highest weight $\lambda_\yn$ of $\rep^\yn$ is also a weight of $\rep^{\yn'}$.
Then $\en(\yn')>\en(\yn)$.
\end{proposition}

\begin{proof}
Denote by $V^\gs_\yn$  the space spanned by the relative ground states of the
$\yn$ multiplet sector $\sptot_\yn$. The corresponding eigenvalue is $\en(\yn)$
according to the notation in Theorem~\ref{tm:main}. So, we have:
$\hamt|_{V^\gs_\yn}\equiv\en(\yn)\cdot\id$
and
$\hamt|_{V^\gs_{\yn'}}\equiv\en(\yn')\cdot\id$.
Note that in case of open chain, according to
Lemma~\ref{lm:multiplet} and Proposition~\ref{prop:same}, both $V^\gs_\yn$ and
$V^\gs_{\yn'}$ are spaces of states
of \emph{irreducible} representations $\rep^\yn$ and $\rep^{\yn'}$ correspondingly.
In case of ring, this is true for
$V^\gs_\yn$ only while the  $\sln$ action on $V^\gs_{\yn'}$ may be \emph{reducible}
consisting of few multiplets each equivalent to $\rep^{\yn'}$.

Consider the weight subspace $\sptot_{N_1\dots N_n}$ corresponding to the highest
weight $\lambda_\yn=\sum_{i=1}^nN_i\eps_i$ of $\rep^\yn$. Remember that as usually we set
$N_i=0$ for all $i>\nu$.
According to Lemma~\ref{lm:subspace} the Hamiltonian $\hamt$ being restricted on it
has nondegenerate ground state $\gs_\yn=\gs_{N_1\dots N_n}$, which belongs to
$V^\gs_\yn$ due to Lemma~\ref{lm:multiplet}. Thus we have:
$\gs_\yn\in \sptot_{N_1\dots N_n}\cap V^\gs_\yn$.
The space $V^\gs_{\yn'}$ has nonzero intersection with the weight subspace $\sptot_{N_1\dots N_n}$
because $\lambda_\yn$ is also a weight of $\rep^{\yn'}$.
Take some state $\gs'_{\yn'}$ belonging to this intersection:
$\gs'_{\yn'} \in \sptot_{N_1\dots N_n}\cap V^\gs_{\yn'}$.
Then  $\left(\gs'_{\yn'},\gs_\yn\right)=0$
because  the subspaces $V^\gs_\yn$ and $V^\gs_{\yn'}$
are mutually orthogonal as non-equivalent $\sln$ representations.
%From the Proposition~\ref{prop:same}
Both states are eigenstates of the Hamiltonian with eigenvalues $\en(\yn)$ and $\en(\yn')$:
$\hamt\,\gs'_{\yn'}=\en(\yn')\,\gs'_{\yn'}$, $\hamt\,\gs_\yn=\en(\yn)\,\gs_\yn$.
Hence, $\en(\yn')>\en(\yn)$ because the relative ground state $\gs_\yn$
of the subspace $\sptot_{N_1\dots N_n}$ is nondegenerate.
This completes the proof.
\end{proof}

Does a simple method exist in order to detect if the highest weight of
some irreducible representation $\rep^\yn$ belongs to the weight space of
another representation $\rep^{\yn'}$?
The problem can be resolved by comparative examination of the shapes
of corresponding Young tableaux $\yn$ and $\yn'$.

\begin{proposition}
\label{prop:ordering-2}
Let $\yn=\yn[N_1,\dots,N_\nu]$ and $\yn'=\yn[N'_1,\dots,N'_{\nu'}]$ be
two different $\sun$ Young tableaux, i.e. $\nu,\nu'\le n$.
Then the weight space of $\rep^{\yn'}$ contains the highest weight of
$\rep^\yn$ if and only if $\yn'\succ\yn$ with
the partial order introduced in Definition~\ref{def:ordering}.
\end{proposition}

\begin{proof}
Recall Definition~\ref{def:young-basis}, where the standard basis for any
$\sln$ representation in terms of Young tableaux is given. To every basic state
a Young tableau, colored according to the following rule, corresponds.
The colors of the boxes along the rows are in nondescending order rightwards,
while along the columns they are in ascending order downward.
See \eqref{bvec} as an example.
%The highest weight vector has the same $i$-th color along the whole $i$-th row
%(see figure~\ref{fig:hvec}).

Suppose that $\yn'\succ\yn$. According to
Definition~\ref{def:ordering}, the tableau $\yn'$ is reduced to
$\yn$ after lowering some of its boxes. In the colored Young
tableau, corresponding to the highest weight state $\Psi_{\yn'}$,
these boxes have the colors equal to their row numbers (see
Fig.~\ref{fig:hvec}). We now relabel them according to their new
row positions without move them down. Obviously, we get in this
way another state of $\rep^{\yn'}$, which also belongs to the
standard basis. It is clear that its weight coincides with the
highest weight $\lambda_\yn$ of $\rep^{\yn}$. This proves the
first part of the proposition. As an example, look at the first
and last tableaux drawn on Fig.~\ref{fig:cmp}.

Suppose now that $\lambda_\yn$ is also a weight of $\rep^{\yn'}$.
%Note that $\lambda_{\yn}\ne \lambda_{\yn'}$ because $\yn'\ncong\yn$.
This means that a state $\Psi^{\lambda_\yn}_{\yn'}$ from the standard basis of $\rep^{\yn}$ exists,
which satisfy the relation $\rep^{\yn}(h)\Psi^{\lambda_\yn}_{\yn'}=\lambda_\yn(h)\Psi^{\lambda_\yn}_{\yn'}$
for any element $h\in\cartan$.
Compare this state with the highest weight state $\Psi_{\yn'}$ of $\rep^{\yn'}$.
According to Definition~\ref{def:young-basis} of the standard basis,
the color of any box in the colored Young tableau corresponding to
$\Psi^{\lambda_\yn}_{\yn'}$ is greater than or equal to its
row position. If it is greater, let us pull down this box to new row position corresponding to
its color. Repeating this procedure for all such boxes, we will arrive to the colored
Young tableau $\yn$ corresponding to the highest weight state $\Psi^{\lambda_\yn}_{\yn}$
of $\rep^\yn$.
Hence, $\yn'\succ\yn$. This proves the second part of the proposition.

\begin{figure}
$$
\Psi_{\yn}={\yc\young(111,22,3)}
\qquad
\qquad
\Psi^{\lambda_{\yn'}}_\yn={\yc\young(114,25,3)}
\qquad
\qquad
\Psi_{\yn'}={\yc\young(11,2,3,4,5)}
$$
\caption{The colored Young tableau corresponding to the highest weight state
$\Psi_{\yn}=\Psi^{\lambda_\yn}_\yn$ of the multiplet $\rep^{\yn}$,
where $\yn=\yn[3,2,1]$ and $\lambda_{\yn}=3\eps_1+2\eps_2+\eps_3$,
is shown on the left.
In the center the same tableau is drawn for the state $\Psi^{\lambda_{\yn'}}_\yn$
of the \emph{same} multiplet corresponding to the weight $\lambda_{\yn'}=2\eps_1+\eps_2+\eps_3+\eps_4+\eps_5$.
The later is the highest weight of another representation $\rep^{\yn'}$, where $\yn'=\yn[2,1,1,1,1]=\yn(5,1)$.
The corresponding highest weight state $\Psi_{\yn'}=\Psi^{\lambda_{\yn'}}_{\yn'}$ is shown on the
right.
}
\label{fig:cmp}
\end{figure}
\end{proof}

So, the proof of the main result of this article is completed now.
Theorem~\ref{tm:main} follows directly from Propositions~\ref{prop:same},
\ref{prop:ordering-1} and \ref{prop:ordering-2}.

\begin{example}
\label{ex:asym}
As an example consider the chain with $N\le n$ sites. This is the simplest case,
where some exacts results exist, which would help to test our results.
It is easy to see that the minimal energy representation
corresponds to unique $N$-th order totally antisymmetric tensor corresponding
to one-column Young tableau $\yn(N)$, i.e. to the \emph{antiferromagnetic} states.
Indeed, only for this representation all adjacent permutations
$\perm_\ind{i\,i+1}$, which make up the Hamiltonian $\hamt$,
take their minimal value, which equals $-1$.
Similarly the \emph{ferromagnetic} states constituting the unique totally symmetric tensor,
which corresponds to one-row Young tableau $\yn[N]$, have the highest energy level,
because the permutations take their maximal value equal $1$ there.
Therefore for any Young tableau $\yn$, which corresponds to other multiplet
$\rep^\yn$ contained in
the $N$-fold tensor product $\rep^N=\otimes^N\rep^\fund$, the inequality
$$
\en(\yn(N))<\en(\yn)<\en(\yn[N]), \quad \text{where} \quad
\en(\yn(N))=\en_{\rm min}=-\sum_{l=1}^{D(N)}J_{l\,l+1}, \quad
\en(\yn[N])=\en_{\rm max}=\sum_{l=1}^{D(N)}J_{l\,l+1}
$$
takes place.
From other side, it is clear that the Young tableau ordering
$\yn(N)\prec\yn\prec\yn[N]$ exists in accordance with Theorem~\ref{tm:main}.
Note that the totally symmetric representation $\rep^{\yn[N]}$ exists for any value of $N$.
Hence the inequality $\en(\yn)<\en(\yn[N])$ for any $\yn\ncong\yn[n]$
is obeyed for $\sun$ symmetric chain of arbitrary length.
\end{example}

\section{Conclusion}
\label{sec:conclusion}
In this article the energy level ordering of $\sun$ symmetric antiferromagnetic
nearest-neighbor chain with fundamental representation is studied.
The translation invariance is not implied and the site-dependent
coupling coefficients can have arbitrary positive values.
Below the main results are summarized briefly and discussed.

Due to $\sun$ symmetry the Hamiltonian is invariant on each $\yn$ multiplet sector,
where the Young tableau $\yn$ parameterizes the equivalence class of $\sun$ multiplets
constituting this sector.
We have proved in this article that for chains with open boundary conditions the
lowest energy level $\en(\yn_1)$ in $\yn_1$ sector is less than the lowest energy
level in $\yn_2$ sector if
Young tableau $\yn_1$ can be obtained from $\yn_2$ pulling some of its boxes
to lower row positions. Using the partial order for Young tableaux introduced
in this article we can write: $\en(\yn_2)<\en(\yn_2)$ if $\yn_1\prec\yn_2$.
Moreover, only one multiplet exists in $\yn_1$ sector,
which has the lowest energy level $\en(\yn_1)$.
In this sense the relative ground state multiplets are nondegenerate.
For chains with periodic boundary conditions the same assertion remains true provided
that all rows of lower Young tableau $\yn_1$ contain either even or odd number of boxes and
there is no restriction on $\yn_2$.

This result generalized well known Lieb-Mattis energy level
ordering theorem \cite{LM62} for $\su{2}$ symmetric models to
$\sun$ symmetric chains. In contrast to $\su{2}$ case, the
ordering is partial for higher symmetries ($n>2$) even for open
chains. This means that not all levels of relative ground states
can be compared in this way. This is clear from the fact that
there are $\sun$ Young tableaux, which can't be obtained from each
other by the displacement of some boxes downward (or upward).
For higher symmetries the method used
in this article is applied to chains with nearest-neighbor
interactions only. Nevertheless, the discovered ordering allows to
obtain a significant information about low energy behavior of
$\sun$ symmetric chains.

As an application of the ordering theorem the $\sun$ structure and degree of degeneracy
of the chain's ground state is found. It was established that the ground state
is $\sun$ singlet and nondegenerate if chain's length $N$ is a multiple of $n$.
This is true both for open chains and for rings.
Alternatively, for open chain, if $N$ divided by $n$ has a nonzero remainder $m$, i.e.
$N\equiv m\pmod{n}$, then the ground state is degenerate with the degree
of degeneracy equal to the binomial coefficient $C_n^m$.
All ground states in this case form $m$-th order
antisymmetric $\sun$ multiplet. In particular, if $m=1$ then it coincides with the
$n$-dimensional fundamental representation.

Another result we have gotten concerns the relative ground states in the weight subspaces.
They have been proved to be unique for open chains. For rings the additional equal parity
condition for all nonzero $\sun$ color multiplicities in the decomposition of corresponding weight
should be fulfilled in order to satisfy the uniqueness condition.
Moreover, if the weight is the highest weight of some $\yn$ representation, when the respective
relative ground state is the highest weight state of the $\yn$ multiplet, which has
the lowest energy value $\en(\yn)$ in entire $\yn$ multiplet sector.

\begin{acknowledgments}
This work is supported by Volkswagen Foundation of Germany, INTAS grant and Swiss SCOPE grant.
\end{acknowledgments}

\appendix*
\section{Perron-Frobenius theorem}
%\label{sec:P-F}

In this appendix for the convenience we formulate and give the proof of
well known Perron-Frobenius theorem (see, for instance, Ref.~\onlinecite{PF}).
First, remember the definition of the connected matrix.

\begin{definition}
\label{def:connectivity}
The matrix $a_{ij}$ is called connected if for any pair $(i,j)$
a sequence of indexes $i_1,i_2,\dots,i_k$ exists
such that $a_{ii_1}a_{i_1i_2}\dots a_{i_kj}\ne 0$.
\end{definition}

\begin{theorem*}[Perron-Frobenius \cite{PF}]
If a connected hermitian matrix $A=\{a_{ij}\}|_{i,j=1}^n$ has no positive
off-diagonal elements then it has nondegenerate ground state $\psi=\{\psi_i\}|_{i=1}^n$
with positive components, i.e. $\psi_i>0$.
\end{theorem*}
\begin{proof}

Denote by $\lambda$ the minimal eigenvalue of $A$ and by $X_\lambda$
the corresponding subspace.
Any vector $\psi$, belonging to $X_\lambda$, obeys
\be
\label{shredinger}
\sum_{i'\ne i}a_{ii'}\psi_{i'}=(\lambda-a_{ii})\psi_i.
\ee
Then the vector $\psi'_i:=|\psi_i|$ with non-negative elements also belongs to this
subspace. This follows from the inequality
\be
\label{mean}
\sum_{i,j}|\psi_i||\psi_j|a_{ij}\le\sum_{i,j}\psi^*_i\psi_ja_{ij}
\ee
and from the properties of $A$ ($a_{ij}\le0$ if $i\ne j$).
Thus, $\psi'$ also satisfies \eqref{shredinger}.

We'll prove now that all elements of $\psi'$ (and $\psi$) are nonzero.
Indeed, suppose that $\psi'_j=0$ for some $j$. From \eqref{shredinger} we get:
$\sum_{i,i\ne j}a_{ji}\psi'_{i}=0$. In order to satisfy this,
the elements $\psi'_i$, for which $a_{ji}\ne0$,  should vanish too.
Because $A$ is a {\em connected} matrix, we conclude that all elements
$\psi'_i$ should vanish. Thus, all $\psi_i$ are nonzero.

Turning back to the inequality \eqref{mean}, we observe now that the equality
takes place only if all coefficients $\psi_i$ have the same sign, otherwise
the left side is {\em strongly} less than the right side. So, if $\psi\in X_\lambda$
then $\psi_i>0$ for all $i=1,\dots,n$ up to an overall sign factor.
This proves the second part of the theorem.

Suppose now, there are two different minimal states $\psi^1,\psi^2 \in X_\lambda$.
From the arguments above, one can consider them to be positive, i.e.
$\psi^{1,2}_i>0$. Any nonzero superposition of $\psi^1$ and $\psi^2$
also belongs to $X_\lambda$. Evidently, among them one can choose one,
which has the coefficients with alternating sign. This contradicts
with the second part of the theorem already proved. Thus, the space $X_\lambda$
is one-dimensional.
\end{proof}

\end{document}

%% file: chain.eepic
\setlength{\unitlength}{0.0007in}
\begingroup\makeatletter\ifx\SetFigFont\undefined%
\gdef\SetFigFont#1#2#3#4#5{%
  \reset@font\fontsize{#1}{#2pt}%
  \fontfamily{#3}\fontseries{#4}\fontshape{#5}%
  \selectfont}%
\fi\endgroup%
{\renewcommand{\dashlinestretch}{30}
\begin{picture}(7857,1196)(0,-10)
\put(3555,71){\blacken\ellipse{128}{128}}
\put(3555,71){\ellipse{128}{128}}
\put(2674,71){\blacken\ellipse{128}{128}}
\put(2674,71){\ellipse{128}{128}}
\put(1035,71){\blacken\ellipse{128}{128}}
\put(1035,71){\ellipse{128}{128}}
\put(154,71){\blacken\ellipse{128}{128}}
\put(154,71){\ellipse{128}{128}}
\put(6435,881){\blacken\ellipse{128}{128}}
\put(6435,881){\ellipse{128}{128}}
\put(5760,881){\blacken\ellipse{128}{128}}
\put(5760,881){\ellipse{128}{128}}
\put(7155,881){\blacken\ellipse{128}{128}}
\put(7155,881){\ellipse{128}{128}}
\put(5130,521){\blacken\ellipse{128}{128}}
\put(5130,521){\ellipse{128}{128}}
\put(7785,521){\blacken\ellipse{128}{128}}
\put(7785,521){\ellipse{128}{128}}
\put(6435,116){\blacken\ellipse{128}{128}}
\put(6435,116){\ellipse{128}{128}}
\put(7155,116){\blacken\ellipse{128}{128}}
\put(7155,116){\ellipse{128}{128}}
\put(5760,116){\blacken\ellipse{128}{128}}
\put(5760,116){\ellipse{128}{128}}
\path(135,71)(1215,71)
\path(135,71)(1215,71)
\dottedline{45}(1260,71)(2475,71)
\path(3555,71)(2475,71)
\path(3555,71)(2475,71)
\drawline(2340,566)(2340,566)
\path(5130,521)(5760,881)(7155,881)
    (7785,521)(7155,116)(5760,116)(5130,521)
\put(45,296){\makebox(0,0)[lb]{\smash{{{\SetFigFont{10}{12}{\rmdefault}{\mddefault}{\updefault}$k_1$}}}}}
\put(400,200){\makebox(0,0)[lb]{\smash{{{\SetFigFont{10}{12}{\rmdefault}{\mddefault}{\updefault}$J_{12}$}}}}}
\put(900,296){\makebox(0,0)[lb]{\smash{{{\SetFigFont{10}{12}{\rmdefault}{\mddefault}{\updefault}$k_2$}}}}}
\put(2350,296){\makebox(0,0)[lb]{\smash{{{\SetFigFont{10}{12}{\rmdefault}{\mddefault}{\updefault}$k_{N-1}$}}}}}
\put(2850,200){\makebox(0,0)[lb]{\smash{{{\SetFigFont{10}{12}{\rmdefault}{\mddefault}{\updefault}$J_{N-1\,N}$}}}}}
\put(3510,296){\makebox(0,0)[lb]{\smash{{{\SetFigFont{10}{12}{\rmdefault}{\mddefault}{\updefault}$k_N$}}}}}
\put(0,971){\makebox(0,0)[lb]{\smash{{{\SetFigFont{14}{16.8}{\familydefault}{\mddefault}{\updefault}(a)}}}}}
\put(4635,1016){\makebox(0,0)[lb]{\smash{{{\SetFigFont{14}{16.8}{\familydefault}{\mddefault}{\updefault}(b)}}}}}
\put(5625,1061){\makebox(0,0)[lb]{\smash{{{\SetFigFont{10}{12}{\familydefault}{\mddefault}{\updefault}$k_1$ }}}}}
\put(5950,990){\makebox(0,0)[lb]{\smash{{{\SetFigFont{10}{12}{\familydefault}{\mddefault}{\updefault}$J_{12}$ }}}}}
\put(6390,1061){\makebox(0,0)[lb]{\smash{{{\SetFigFont{10}{12}{\familydefault}{\mddefault}{\updefault}$k_2$}}}}}
\put(6650,990){\makebox(0,0)[lb]{\smash{{{\SetFigFont{10}{12}{\familydefault}{\mddefault}{\updefault}$J_{23}$ }}}}}
\put(7155,1061){\makebox(0,0)[lb]{\smash{{{\SetFigFont{10}{12}{\familydefault}{\mddefault}{\updefault}$k_3$}}}}}
\put(7415,820){\makebox(0,0)[lb]{\smash{{{\SetFigFont{10}{12}{\familydefault}{\mddefault}{\updefault}$J_{34}$}}}}}
\put(7785,700){\makebox(0,0)[lb]{\smash{{{\SetFigFont{10}{12}{\familydefault}{\mddefault}{\updefault}$k_4$}}}}}
\put(7470,200){\makebox(0,0)[lb]{\smash{{{\SetFigFont{10}{12}{\familydefault}{\mddefault}{\updefault}$J_{45}$}}}}}
\put(7000,296){\makebox(0,0)[lb]{\smash{{{\SetFigFont{10}{12}{\familydefault}{\mddefault}{\updefault}$k_5$}}}}}
\put(6650,230){\makebox(0,0)[lb]{\smash{{{\SetFigFont{10}{12}{\familydefault}{\mddefault}{\updefault}$J_{56}$ }}}}}
\put(6345,296){\makebox(0,0)[lb]{\smash{{{\SetFigFont{10}{12}{\familydefault}{\mddefault}{\updefault}$k_6$}}}}}
\put(5950,230){\makebox(0,0)[lb]{\smash{{{\SetFigFont{10}{12}{\familydefault}{\mddefault}{\updefault}$J_{67}$ }}}}}
\put(5715,296){\makebox(0,0)[lb]{\smash{{{\SetFigFont{10}{12}{\familydefault}{\mddefault}{\updefault}$k_7$}}}}}
\put(5150,200){\makebox(0,0)[lb]{\smash{{{\SetFigFont{10}{12}{\familydefault}{\mddefault}{\updefault}$J_{78}$}}}}}
\put(4905,700){\makebox(0,0)[lb]{\smash{{{\SetFigFont{10}{12}{\familydefault}{\mddefault}{\updefault}$k_8$}}}}}
\put(5250,850){\makebox(0,0)[lb]{\smash{{{\SetFigFont{10}{12}{\familydefault}{\mddefault}{\updefault}$J_{18}$}}}}}
\end{picture}
}